\documentclass{article}%
\usepackage{amsmath}
\usepackage{graphicx}%
\usepackage{amsfonts}%
\usepackage{amssymb}

\begin{document}

\title{\emph{Information Processing in Brain Microtubules}}
\author{\textit{Jean Faber}$^{1}$\textit{, Renato Portugal}$^{1}$\textit{, Luiz
Pinguelli Rosa}$^{2}$\\$^{1}${\small Laborat\'{o}rio Nacional de Computa\c{c}\~{a}o Cient\'{\i}fica -
LNCC, }\\{\small Av. Get\'{u}lio Vargas 333 - Quitandinha, 25651-075,}\\{\small Petr\'{o}polis, RJ, Brazil.}\\{\small \{faber, portugal\}@lncc.br}\\$^{2}${\small Universidade Federal do Rio de Janeiro, }\\{\small COPPE-UFRJ, RJ, Brazil.}\\{\small lpr@adc.coppe.ufrj.br}}
\maketitle

\begin{abstract}
Models of the mind are based on the idea that neuron microtubules can perform
computation. From this point of view, information processing is the
fundamental issue for understanding the brain mechanisms that produce
consciousness. The cytoskeleton polymers could store and process information
through their dynamic coupling mediated by mechanical energy. We analyze the
problem of information transfer and storage in brain microtubules, considering
them as a communication channel. We discuss the implications of assuming that
consciousness is generated by the subneuronal process.\bigskip

\end{abstract}

\section{Introduction}

\qquad In recent years many papers have addressed the problem of developing a
theory of mind [1-13]. R. Penrose and S. Hameroff developed a quantum model of
the mind considering the cytoskeleton of neuron cells as the principal
component that produces states of mind or consciousness [2,3]. In their model
the microtubules (MTs) perform a kind of quantum computation through the
tubulins. Tubulins are proteins which form the walls of the MTs. They claim
that the tubulins work like a cellular automata performing that kind of
computation. In this way, the walls of the MT could be able to store and
process information by using combinations of the two possible states ($\alpha$
and $\beta$) of the tubulins. The MT interior works as an electromagnetic wave
guide, filled with water in an organized collective state, transmitting
information through the brain. A gelatinous state of water in brain cells,
which was observed by [13], could boost these communication effects.

Using a different approach, Tuszynski et al. [6-8] model the biophysical
aspects of the MTs considering the following questions: What kind of computing
do microtubules perform? How does a microtubule store and process information?
In order to analyze these questions they use a classical approach, studying
the basic physical properties of the MTs as interacting electric dipoles.

According to [6-8,14-17] each tubulin has an electric dipole moment
$\overrightarrow{p}$ due to an asymmetric charge distribution. The microtubule
is thus a lattice of oriented dipoles that can be in random phase,
ferroelectric (parallel-aligned) and an intermediate weakly ferroelectric
phase like a spin-glass phase. It is natural to consider the electric field of
each tubulin as the information transport medium.

Therefore, the tubulin dimers would be considered the information unit in the
brain and the MT sub-processors of the neuron cells. Therefore, to know how
MTs process information and allow communication inside the brain is a
fundamental point to understand the mind functions.

In this work we derive some results which were not explicitly obtained in
[6-8] and extend the ideas introduced by [6,16] using the point of view of the
information theory. We analyze the problem of information transfer and storage
in brain microtubules, considering them as a communication channel. The
electric field is the mediator of each communicator entity. We discuss the
implications of assuming that the consciousness is generated by the
microtubules as sub-neuronal processors.

\section{Biophysical Aspects of the Microtubules}

\qquad The cytoskeleton has a dynamic structure which reorganizes continually
as the cells change their shape, divide, and respond to their environment. The
cytoskeleton is composed of intermediate filaments, actin filaments (or
microfilaments), and microtubules. The filaments and the microtubules are
mutually connected and form a three-dimensional network in the cell. There are
many papers [6-8] showing that the cytoskeleton is the main component which
organizes the cell, mediates transport of molecules, organelles, and synaptic
vesicles. The cytoskeleton possibly receives signals from the cellular
environment mediated by the membrane of proteins and participates in signal
transmission to the neighborhood of the cell [16,17].

Microtubules are hollow cylinders whose exterior surface cross-section
diameter measures $25nm$ with $13$ arrays of protein dimers called tubulins.
The interior of the cylinder contains ordered water molecules which implies
the existence of an electric dipole moment and an electric field. The MTs
represent a dipole due to individual dipolar charges of each tubulin monomer.
The microtubule dipole produces a fast growth at the plus end towards the cell
periphery and a slow growth at the minus end. The MT polarity is closely
connected with its functional behavior which can be regulated by
phosphorylation and dephosphorylation of microtubule-associated protein (MAP) [6-8,14-17].

Guanosine triphosphate molecules (GTP) are bound to both tubulins in the
heterodimer. After polymerization, when the heterodimer is attached to the
microtubule, the GTP bound to the $\beta$-tubulin is hydrolyzed to the
guanosine disphosphate (GDP). On the other hand, the GTP molecule of the
$\alpha$-tubulin is not hydrolyzed. The microtubules present a calm dynamic
instability which are their principal feature [6-8].\cite{Sari:2000:ATAB}

Many models of conformation (and polarity energy) of the microtubular
protofilament were developed. These models describe the behavior of the pulses
generated by the free energy in the GTP hydrolysis. The pulses propagate along
of the MTs through an elastic coupling or through electric field propagation
between tubulin dimers [5-8,14,15]. The overall effect of the surrounding
dipoles on a site $n$ can be modelled by the double-well quartic potential [6-7]%

\begin{equation}
V(u_{n})=-\frac{A}{2}u_{n}^{2}+\frac{B}{4}u_{n}^{4},\label{Hamilt}%
\end{equation}
where $u_{n}$ represents the dimer conformational change on the \textit{n--th}
protofilament axis coupled to the dipole moment. $A$ and $B$ are parameters of
the model, where $A$ is dependent of the temperature by $\ A=a(T-T_{c})$,
$T_{c}$ is the critical temperature, and $B$ is a positive parameter
independent of the temperature [6,8]. In figure 1 we plot the effective
potential in terms of $u_{n}$.%
\[%
{\parbox[b]{1.6518in}{\begin{center}
\includegraphics[
natheight=1.791900in,
natwidth=2.052200in,
height=1.2678in,
width=1.6518in
]%
{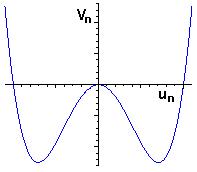}%
\\
{\small Fig. 1 - Double well quartic potential model with a potential barrier
}$|A^2/2B|.$
\end{center}}}%
\]

Our assumptions lead us to reconsider this model taking into account the
Information Theory to calculate the storage and transference of information
along the MT. The information is mediated by the electric field propagating in
the cellular medium. This propagation of energy can provide a communication channel.

\section{Communication Channels}

\qquad The Shannon entropy of a random variable $X$ is defined by [18]:%
\begin{equation}
\langle I(X)\rangle=-\sum_{i}p(x_{i})\log p(x_{i}). \label{Info}%
\end{equation}
where $p(x_{i})$ is the probability of the outcome $x_{i}.$ This definition
describes the amount of physical resources required on average to store the
information being produced by a source, in such a way that at a later time the
information can be restored completely.

If we want to send a message $X$ through a noisy channel, that message can be
subjected to a loss of information. To correlate a sent message $X$ with a
received message $Y$ we have to calculate the \textit{mutual information}
$I(X:Y)$ between them. The \textit{mutual information} concept gives us how
much knowledge we obtain from a message $X$ given that we have received $Y.$
It is defined by [18,19]%
\begin{equation}
\langle I(X:Y)\rangle=\langle I(X)\rangle-\langle I(X|Y)\rangle=\langle
I(Y)\rangle-\langle I(Y|X)\rangle\label{mutual info}%
\end{equation}
and
\begin{equation}
\langle I(X|Y)\rangle=-\sum_{i}\sum_{k}p(x_{i},y_{k})\log p(x_{i}|y_{k}),
\label{join info}%
\end{equation}
where $p(x_{i}|y_{k})=p(y_{k},x_{i})/p(y_{k}).$

Nevertheless, by using a binary code to send a message $M,$ compressed by
procedure $C$ that minimizes the use of bits in that codification, any
receiver of $M,$ using a decoding procedure $D,$ must to be able to get all
information associated to $M.$

Consider a symmetric memoryless channel\footnote{The memoryless channel is the
one that acts in the same way every time it is used, and different uses are
independent of one another.} $N$ with a binary input $A_{in}$ and a binary
output $A_{out}$. For $n$ uses of the channel, the procedure $C$ encodes the
input message $M$ such that $C^{n}:\{1,...,2^{nR}\}\rightarrow A_{in}$ and $D$
decodes the output such that $D^{n}:\{1,...,2^{nR}\}\rightarrow A_{out},$
where $R$ is the \textit{rate of the code }(the\textit{ }number of data bits
carried per bit transmitted) [19]. Therefore, if $X$ is the encoded message
$M$ through the procedure $C,$ $Y$ is the received message, and $D$ is the
decoding procedure for $Y$, then the probability of error is defined by%
\begin{equation}
p(C^{n}.D^{n})=\max_{M}p(D^{n}(Y)\neq M|X=C^{n}(M)). \label{prob erro}%
\end{equation}

The principal problem of the information theory is to determine the maximum
rate $R$ for a reliable communication through a channel. When $p(C^{n}.D^{n})$
$\rightarrow$ $0$ for $n\rightarrow$ $\infty,$ the rate $R$ is said
achievable. According to Shannon's theorem, given a noisy channel $N,$ its
capacity $\Omega(N)$ is defined to be the supremum over all achievable rates
for this channel. That is%
\begin{equation}
\Omega(N)=\max_{p(x_{i})}(\langle I(X:Y)\rangle), \label{canal}%
\end{equation}
where the maximum is taken over all input distributions $p(x_{i})$ of the
random variable $X$, for one use of the channel, and $Y$ is the corresponding
induced random variable at the output of the channel.

Equation (\ref{canal}) allows us to calculate the transference of information
among many physical systems. The transfer of energy may include the transfer
of electrostatic energy, energy of low frequency oscillating fields, energy of
light, energy of vibrations, etc. Molecules can contain energy in the chemical
bonds, in the excited electron states, in the conformation states, etc. A
common measure of the interaction leading to cooperative behaviour is the
information transference. The electromagnetic field can transfer information
through the environment among the systems like a communication channel.

\section{Information Processing in Microtubules}

\qquad Many features of the cytoskeleton support the idea that microtubules
can perform computation and store information. According to [6] the charge
separation of the MTs is wide enough to store information. Due to its dynamic
coupling the information can be stored as mechanical energy and chemical events.

Changes in the opposite direction can be favorable to the SG phase over the
F-phase. This change could switch from the growth mode to operational
behavior. Our focus is this operational mode. Information processing is
addressed by [1-5] considering the highly specialized nature of the functional
proteins on the microtubules.

\subsection{Information Storage in Microtubules}

\qquad The tubulins form a dipole moment net and therefore are sensitive to
external electric fields. Some papers use physical models such as spin net to
describe the behavior of the dipole moment net [6-7,20]. According to those
models models, all tubulins are oriented to the same direction at low
temperature ($\sim200K$) and the units of the system are organized (figure 2).
In this case the system is in ferroeletric phase (F). At high temperatures
($\sim$ $400K$), the system is in the paraelectric phase (P) and the polarity
of the tubulins are completely disorganized (figure 3). However, there is a
critical temperature $T_{c}$ in which occurs a phase transition between F and
P, that is, between order and disorder. At this phase transition emerges a new
state known as spin-glass phase (SG) (figure 4). There are some theoretical
models trying to estimate this critical temperature. One of them estimates the
critical temperature around to $300K$ which is near to the human body
temperature [7,8].%
\[%
{\parbox[b]{1.8792in}{\begin{center}
\includegraphics[
natheight=0.614900in,
natwidth=1.760800in,
height=0.6746in,
width=1.8792in
]%
{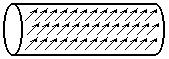}%
\\
{\small Fig. 2 - schematic picture for F-phase}%
\end{center}}}%
\]%
\[%
{\parbox[b]{1.868in}{\begin{center}
\includegraphics[
natheight=0.604500in,
natwidth=1.750400in,
height=0.6633in,
width=1.868in
]%
{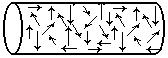}%
\\
{\small Fig. 3 - schematic picture for P-phase}%
\end{center}}}%
\]%
\[%
{\parbox[b]{1.8792in}{\begin{center}
\includegraphics[
natheight=0.635600in,
natwidth=1.760800in,
height=0.6962in,
width=1.8792in
]%
{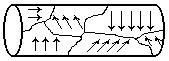}%
\\
{\small Fig. 4 - schematic picture for SG-phase}%
\end{center}}}%
\]

We analyze the propagation of informatin along MTs considering the above
phases. Assuming an energy approximation dependent on the mean polarity
described by the Landau theory of phase transitions, the total energy can be
given by [6,16,21]%
\begin{equation}
E=\left(  \frac{a}{2}\wp^{2}+\frac{b}{4}\wp^{4}\right)  N_{0}, \label{energy}%
\end{equation}
where $\wp$ represents the continuous variable for the mean polarization at
each site, and $N_{0}$ is the total number of sites. The parameter $a$ has a
linear dependence with the temperature $a=\overline{a}(T-T_{c})$, where
$200K<T_{c}<$ $400K$ and $b>0$ [6-8]. $E$ will be minimized by $\wp=0$ for
$T>T_{c}$ and by $\wp=\pm\sqrt{-\overline{a}(T-T_{c})/b}$ for $T<T_{c}.$ We
use the Boltzmann distribution $g(\wp)$ to weight the energy distribution as a
function of the mean polarity%
\begin{equation}
g(\wp)=Z^{-1}\exp(-\beta E), \label{MB distribution}%
\end{equation}
where $\beta^{-1}=kT$, $Z$ is the normalization, and $k$ is the Boltzmann
constant. Subtituting (\ref{MB distribution}) into (\ref{energy}) we get%
\begin{equation}
g(\wp)=Z^{-1}\exp\left(  \frac{\overline{a}(T-T_{c})}{2kT}\wp^{2}+\frac
{b}{4kT}\wp^{4}\right)  . \label{dist applied}%
\end{equation}

Because $\wp$ is a continuous variable, we need to use the continuous
counterpart of (\ref{Info}) in order to calculate the information mean value
of the system. Replacing $p(x)$ by $g(\wp)$ in (\ref{Info}) we obtain the
following expression for the information storage capacity:
\begin{equation}
\langle I\rangle=\frac{\ln Z}{\ln2}-\frac{\overline{a}(T-T_{c})}{2kT\ln
2}\langle\wp^{2}\rangle-\frac{b}{4kT\ln2}\langle\wp^{4}\rangle.
\label{info capacity}%
\end{equation}
The average of $\wp$ over the whole MT, considering all domains, is obtained
from%
\begin{equation}
\langle\wp^{n}\rangle=\int_{-\infty}^{\infty}g(\wp)\wp^{n}d\wp.
\label{polarity mean}%
\end{equation}

Using (\ref{info capacity}) we can plot the information capacity against the
temperature for some parameter values.%
\[%
{\parbox[b]{3.0623in}{\begin{center}
\includegraphics[
natheight=2.166400in,
natwidth=2.885000in,
height=2.3065in,
width=3.0623in
]%
{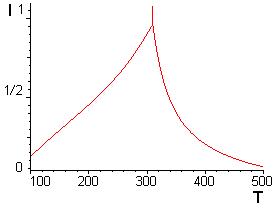}%
\\
{\small Fig 5. - Storage information capacity of MT when }$\overline
{a}=b=0.5.$
\end{center}}}%
\]%
\[%
{\parbox[b]{3.0952in}{\begin{center}
\includegraphics[
natheight=2.260600in,
natwidth=2.917000in,
height=2.405in,
width=3.0952in
]%
{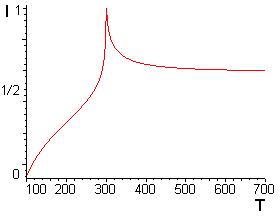}%
\\
{\small Fig. 6 - Storage information capacity of MT when }$\overline
{a}{\small =}0.5$ and $b=50.$
\end{center}}}%
\]%
\[%
{\parbox[b]{3.0735in}{\begin{center}
\includegraphics[
natheight=2.187100in,
natwidth=2.896300in,
height=2.3281in,
width=3.0735in
]%
{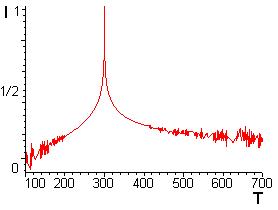}%
\\
{\small Fig. 7 - Storage information capacity of MT when }$\overline
{a}\;{\small =\;}0.05$ and $b=50.$
\end{center}}}%
\]

These graphs corroborate with the results of [6], which show that, probably at
physiological temperature, we can have a mode of information storage in MTs.
This is the most important feature for finding another subunit of information
processing inside the brain. It could show us new perspectives for cognitive aspects.

However, according to these graphs, the maximum information storage is
obtained at the spin-glass phase, therefore we need to make some assumptions.
In this phase, there are domains with many energy levels which can store
information. The interaction among the domains due to the electric field
generated by the oscillating dipoles must be considered. This electric field
is emitted to the neighbouring area producing many channels among the domains
in MT.

\subsection{Microtubules as a Communication Channel}

\qquad Given the capacity of information storage of MTs, the issue now is to
know whether there exist some kind of information processing on them. To study
any kind of processing, it is necessary to describe how the information is
stored in the MT walls. That is, we need to understand how the information
propagates along the MT. We saw that the SG phase has the maximum capacity of
information storage. Therefore, we will restrict to this phase in order to
describe the interaction, or communication, among the domains. Here, we are
assuming that the electric field generated by the MT dipoles is the main
mediator which allows the communication among the domains.

The graphs of the previous section show that near to the critical temperature
$T_{c}$ the information capacity has the maximum capacity of storage.
Following [6], we assume in this phase a partition of lattices by local
domains (see figure 4). Therefore, the previous prescription is valid only on
the local domains. In this way, a domain $j$ has a polarization $\wp_{j}$ with
probability $g_{j}(\wp_{j})$. If we make these assumptions, the total
probability is given by%
\begin{equation}
g=\prod_{j=1}^{r}g_{j}(\wp_{j}), \label{partition distribution}%
\end{equation}
where $r$ is the number of domains [6]

As a consequence of (\ref{partition distribution}) we have for the spin-glass
phase%
\begin{equation}
\langle I\rangle=\sum_{j}\langle I_{j}\rangle\label{info sum}%
\end{equation}
with $j$ in the set of domains.

Now, we need to calculate the amount of information transferred through the
channels among the domains [6,16]. The domains will communicate only if they
interact. If we consider two domains, the communication is mediated by the
electric field interaction between them. In order, to calculate the capacity
of this communication channel, we use the mutual information concept.

From (\ref{canal}) we know how much information is transferred from an event
$x_{k}$ (of an ensemble $X)$ to another event $y_{j}$ (of an ensemble $Y).$
The term $p(x_{k}|y_{j})$ imposes the dependence among the systems. Assuming a
Boltzmann distribution, we want to know the dependence between the
\textit{domain }$k$, with polarization $\wp_{k},$ and the \textit{domain }$j$,
with polarization $\wp_{j}.$ This dependence is described by the distribution
$g(\wp_{k}|\wp_{j})$ which imposes a connection between the domains. Following
[16] we will assume that\ the output energy is expressed as a function of the
electric field energy and of the mean polarization energy. Therefore, in the
thermodynamic equilibrium, we have the average of the output energy $E^{out}$
of a \textit{domain }$j$\textit{ }as%
\begin{equation}
\langle E_{j}^{out}\rangle=\langle E_{j}^{signal}\rangle=\langle E_{j}%
^{flow}\rangle+\langle E_{j}^{noisy}\rangle, \label{output energy}%
\end{equation}
where $E_{j}^{signal}$ is the energy of the coherent signal, $E_{j}^{noisy}$
is the noisy energy, and $E_{j}^{flow}$ is the energy of the flow along the
system. The energy $E_{j}^{flow}$ is responsible for the interaction between
the domains. Therefore, supposing that the \textit{domain }$j$ emitts
$E_{j}^{flow}$, we can express the dependence of a \textit{domain }$k$ as%
\begin{equation}
g(\wp_{k}|\wp_{j})=Z^{-1}\exp\left[  -\beta(E_{k}+E_{j}^{flow}+E_{j}%
^{noise})\right]  , \label{mutual distribution}%
\end{equation}
where $E_{k}$ is the correspondent energy of the \textit{domain }$k$.

The information entropy depends on the amount of energy in the system and on
the noisy energy $E_{j}^{noisy}.$ Therefore, the energy of the noise is given
by [16]%
\begin{equation}
\langle E_{j}^{noisy}(T_{n})\rangle=Z^{-1}\exp(A\langle\wp_{j}^{2}%
\rangle+B\langle\wp_{j}^{4}\rangle), \label{noisy energy}%
\end{equation}
where$%
\begin{array}
[c]{c}%
A=\overline{a}(T_{n}-T_{c})/2kT_{n}%
\end{array}
$ and $%
\begin{array}
[c]{c}%
B=b/4kT_{n}%
\end{array}
.$

To evaluate the communication channel capacity, each domain is approximated by
a unique dipole. The information transference will be mediated by a radiation
of the electric field in the equatorial region of an oscillating dipole.

Using the complex Poynting vector and taking the real part, we get an
expression for the mean value of the flow of energy $E_{j}^{flow}$. The amount
of energy absorbed by the oscillating charged units depends directly on its
effective cross-section and on the intensity of the flow of energy. We can
calculate it considering the radiation flow of energy through the
cross-section $D$ over a spherical surface of radius $R,$ where $D$ is a
rectangle whose sides are $x$ and $z$. Hence the expression for the flow of
energy towards the dipole axis is given by%
\begin{equation}
\langle E_{j}^{flow}\rangle=2\pi^{2}S_{j}\arcsin\frac{x}{2R_{x}}\left[
\frac{z}{2R_{z}}-\frac{1}{3}\left(  \frac{z}{2R_{z}}\right)  ^{3}\right]  ,
\label{elelctric field}%
\end{equation}
where $S_{j}=\wp_{j}^{2}\sqrt{\eta^{3}\varepsilon}\nu_{j}^{4},$ such that
$\nu_{j}$ is the dipole frequency, $\varepsilon$ is the permitivity, $\eta$ is
the permeability of the medium, $R_{x}$ and $R_{z}$ are the perpendicular
distances from the dipole to the $z$ and $x$ sides, respectively [16].

Using (\ref{mutual info}) and (\ref{join info}), and the previous relations,
we can derive an expression for the capacity of communication between two
domains. The channel between two domains $j$ and $k$ will be denoted by
$N_{jk},$ hence%
\begin{equation}
\Omega(N_{jk})=\langle I(E_{k})\rangle-\langle I(E_{k}|E_{j})\rangle,
\label{domain channel}%
\end{equation}
where $\langle I(E_{k})\rangle$ is given by an expression similar to
(\ref{info capacity}). The conditional information $\langle I(E_{k}%
|E_{j})\rangle$, for a specific polarity $\wp_{j},$ can be calculated by a
continuous version of (\ref{join info}), that is%
\begin{equation}
\langle I(E_{k}|E_{j})\rangle=\frac{\ln Z}{\ln2}-\beta\int g(\wp_{k},\wp
_{j})\left(  E_{k}+E_{j}^{signal}\right)  d\wp_{k}. \label{domain cond inf}%
\end{equation}

Through those calculations we can infer that there is an inter-dependence
among the domains in the SG phase. Each domain communicates to other domain
the value of its polarity. It transforms the MT in a net of communication
units (in this case the units are the domains - see figure 8). Besides, as
each domain has a particular polarity, in the context of the information
theory, we can interpret each polarity representing a type of symbol. It would
build a kind of alphabet along the whole MT, where each domain represents a
letter. However, as the polarity $\wp$ is a continuous variable, the change of
a letter in another would be also in a continuous way and not in a discrete
way%
\[%
\raisebox{-0cm}{\parbox[b]{5.1906cm}{\begin{center}
\includegraphics[
trim=0.000000in 0.000000in -0.037671in -0.009086in,
natheight=1.781500in,
natwidth=4.280800in,
height=2.1725cm,
width=5.1906cm
]%
{communMT.bmp}%
\\
{\small Fig. 8 - A representation of the communication between domain}
${\small j}$ {\small and domain} ${\small k}$ {\small accomplished by the
electromagnetic field }${\small E}_j^flow${\small along the walls of MT}.
\end{center}}}%
\]

Considering the case $x=z,$ we plot the capacity of information transference
between each domain as a function of the distance and frequency.%
\[%
{\parbox[b]{2.7665in}{\begin{center}
\includegraphics[
natheight=2.052200in,
natwidth=2.603900in,
height=2.1854in,
width=2.7665in
]%
{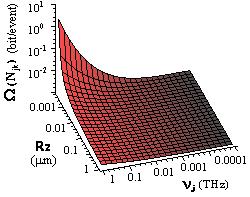}%
\\
{\small Fig. 9} - {\small Communication channel capacity: frequency }$%
\upsilon_j\times$ distance $R_z$ when ${\small T\sim300K.}$
\end{center}}}%
\]%
\[%
{\parbox[b]{2.7337in}{\begin{center}
\includegraphics[
natheight=2.073000in,
natwidth=2.572800in,
height=2.2087in,
width=2.7337in
]%
{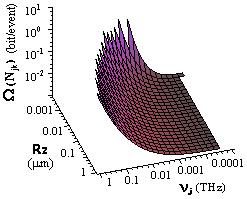}%
\\
{\small Fig. 10} - {\small Communication channel capacity: frequency }%
$\upsilon_j\times$ distance $R_z$ when ${\small T\sim100K.}$
\end{center}}}%
\]%
\[%
{\parbox[b]{2.7233in}{\begin{center}
\includegraphics[
natheight=2.083300in,
natwidth=2.562400in,
height=2.2191in,
width=2.7233in
]%
{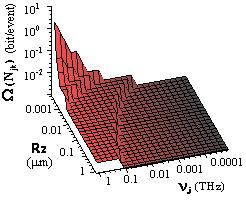}%
\\
{\small Fig. 11} - {\small Communication channel capacity: frequency }%
$\upsilon_j\times$ distance $R_z$ when ${\small T\sim600K.}$
\end{center}}}%
\]

The graphs show that the best conditions to have a communication among the
domains are at the physiological temperature, with frequency of the
conformational changes of the tubulin dimer protein around to 10$^{12}s^{-1}$.
The relative permitivity and permeability in the neighbourhood of the
oscillating units is assumed to be $1$ [14-16]. The distance $z$ between the
protein molecules is adopted to be around to $1%
\mu
m-0.1%
\mu
m$. At $300K$ the information transference is supressed over a distance
$R_{z}$ equal to $0.1%
\mu
m$, and frequency around to $0.1THz$ (figure 9). At a distance smaller than
$0.1%
\mu
m$ the communication starts to become independent of the frequency. Finally,
at a distance greater than$0.1%
\mu
m$ the high frequency of the electric field plays a fundamental role in the
transfer of information. For the other regimes of temperature, the system is
not in the SG-phase and the graphs show the loss in performance (figures 10
and 11).

According to [22], biological molecules with dipolar vibrational activity
could manifest a quantum coherent mode. That systems could have some isolating
effect from thermal environments. The frequency range of that quantum mode,
(also known as Fr\"{o}hlich systems) is around $10^{11}s^{-1}$ to
$10^{12}s^{-1}$ [1,4]. Therefore, the high frequency regimes obtained above
can work not only to perform a communication along the MT but also to maintain
some quantum coherent regime\footnote{Some papers show that the tubulin
vibration frequency is in this regime [16,21].}.

\section{Conclusions}

\qquad This work confirm the results of [1-8] which consider microtubules as a
classical subneuronal information processor. Through the information theory we
calculate the information capacity of the MTs. Utilizing models of [1-8] we
estimate that the favorable conditions for storage and information processing
are found at temperatures close to the human body. These results corroborate
the possibility of communication among the domains (where each energy level
corresponds to some kind of symbol). This communication is mediated by the
dipole electric field, and this interaction is necessary to describe some
processing or computing on MT. Through this communication, each domain (or
symbol) presents some dependence with another. Therefore there are storage as
well as processing of information associated to the dimers. Besides, from the
information theory point of view, the formation of domains creates some
redundancy for storage or representation of these symbols. This redundancy is
important for error correction and information protection. However, some
points still need further investigations. To mention at least two, (1) the
direction of the propagation of the information under the influence of the
environment is an interesting point to be analyzed, (2) according to [1-5]
there is some water ordination inside MTs which could increase the quantum
processes in MTs. These points deserve to be analyzed using the information
theory point of view.

\section{References}

[1] S. Hagan, S. R. Hameroff and J. A Tuszynski, Quantum Computation in Brain
Microtubules: Decoherence and Biological Feasibility, Phys. Rev. E. 65,
061901, 2002.

\noindent\lbrack3] S. R. Hameroff and R. Penrose, Orchestrated Reduction of
Quantum Coherence in Brain Microtubules, in S. Hameroff, A . K. Kasszniak and
A .C. Scott, Toward a Sience of Consciousness, MIT Press, Cambridge, 1996

\noindent\lbrack3] S. R. Hameroff and R. Penrose, Conscious Events as
Orchestrated Space Time Selections, in J. Shear, \textit{Explaining
Consciousness. The Hard Problem}, MIT Press, Cambridge, USA, 1998

\noindent\lbrack4] M. Jibu, S. Hagan, S. R. Hameroff, K. H. Pribram and K.
Yasue, Quantum Optical Coherence in Cytoskeletal Microtubules: Implications
for Brain Function, Biosystems 32, 195-209, 1994.

\noindent\lbrack5] L. P. Rosa and J. Faber, Quantum Models of Mind: Are They
Compatible with Environmental Decoherence? Phys. Rev. E 70, 031902, 2004.

\noindent\lbrack6] J. A. Tuszynski, B. Tripisova, D. Sept, M.V. Sataric, The
Enigma of Microtubules and their Self-organization Behavior in the
Cytoskeleton. Biosystems 42, 153-175, 1997.

\noindent\lbrack7] J. A. Tuszynski, J. A. Brown and P. Hawrylak, Dieletric
Polarization, Electric Conduction, Information Processing and Quantum
Computation in Microtubules. Are They Plausible?, The Royal Society,356, 1897-1926,1998.

\noindent\lbrack8] J.A. Tuszynski, S. R. Hameroff, M.V. Sataric, B. Trpisova
and M. L. A. Nip, Ferroelectric Behavior in Microtubule Dipole Lattices:
Implications for Information Processing, Signaling and
Assembly/Disassembly.Journal Theoretical Biology, 174, 371-380, 1995.

\noindent\lbrack9] M. Tegmark, Importance of Quantum Decoherence in brain
process, Phys. Rev. E 61 , 4194, 2000

\noindent\lbrack10] R. Penrose, \textit{The Emperor New Mind}, Oxford
University Press, 1989

\noindent\lbrack11] R. Penrose, \textit{Shadows of the Mind}, Vintage, London, 1994

\noindent\lbrack12] J. C. Eccles, \textit{How the Self Controls Its Brain},
Spring Verlag, Berlin, 1994

\noindent\lbrack13] J. Watterson, Water Clusters: Pixels of Life, in S.
Hameroff et alli, Toward a Sience of Consciousness, MIT Press, Cambridge, 1996

\noindent\lbrack14] M. V. Sataric, J. A. Tuszyski and R. B. Zakala Phys. Rev.
E, 48, 589,1993

\noindent\lbrack15] J. A. Brown and J. A. Tuszynski, Phys. Rev. E 56, 5834,1997

\noindent\lbrack16] J. Pokoorny and T. Ming Wu, \textit{Biophysics Aspects of
Coherence and Biological Order}, Spinger, 1998.

\noindent\lbrack17] E. R. Kandel, J. H. Schwarts, and T. M. Jessell,
\textit{Principles of Neural Science.} Appleton \& Lange Norwalk, third
edition 1991.

\noindent\lbrack18] M. A. Nielsen and I. L. Chuang, \textit{Quantum Computing
and Quantum Information}. Cambridge, 2000.

\noindent\lbrack19] J. Preskill, \textit{Lecture notes for Physics 229:
Quantum Information and Computation}, 1998. www.theory.caltech.edu/$\sim$preskill/ph229.

\noindent\lbrack20] V. Dotsenko, An Introduction to The Theory of Spin Glasses
and Neural Networks. World Scientific Lecture Notes in Physics, v. 54, 1994.

\noindent\lbrack21] H. Haken, \textit{Synergetics: An Introduction. }Springer,
Berlin 1990.

\noindent\lbrack22] H. Frohlich, \textit{Coherent excitations in active
Biological systems, in Modern Bioelectrochemistry}, F. Gutman and H. Keyzer,
Springer Verlag, N Y, 1986
\end{document}